\documentclass{emulateapj}
\pdfoutput=1

\shorttitle{A planetary nebula around nova V458 Vul}
\shortauthors{R.\ Wesson et al.}

\begin{document}

\title{A planetary nebula around nova V458 Vul undergoing flash
ionization}

\author{R.\ Wesson\altaffilmark{1}, M. J. Barlow\altaffilmark{1}, R. L. M. Corradi\altaffilmark{2,3}, J. E. Drew\altaffilmark{4}, P. J. Groot\altaffilmark{5}, C. Knigge\altaffilmark{6}, D. Steeghs\altaffilmark{7,10}, B. T. Gaensicke\altaffilmark{7}, R. Napiwotzki\altaffilmark{4}, P. Rodriguez-Gil\altaffilmark{2,3}, A. A. Zijlstra\altaffilmark{8}, M. F. Bode\altaffilmark{9}, J. J. Drake\altaffilmark{10}, D. J. Frew\altaffilmark{11}, E. A. Gonzalez-Solares\altaffilmark{12}, R. Greimel \altaffilmark{13}, M. J. Irwin\altaffilmark{12}, L. Morales-Rueda\altaffilmark{5}, G. Nelemans\altaffilmark{5}, Q. A. Parker\altaffilmark{11}, S. E. Sale\altaffilmark{14}, J. L. Sokoloski\altaffilmark{15}, A. Somero\altaffilmark{16,17}, H. Uthas\altaffilmark{6,16}, N. A. Walton\altaffilmark{12}, B. Warner\altaffilmark{6,18}, C. A. Watson\altaffilmark{19}, N. J. Wright\altaffilmark{1,10}}

\altaffiltext{1}{Department of Physics and Astronomy, University College London, Gower Street, London, WC1E 6BT, UK}
\altaffiltext{2}{Isaac Newton Group, PO Ap. de Correos 321, 38700 Sta. Cruz de la Palma, Spain}
\altaffiltext{3}{Instituto de Astrof{\'{\i}}sica de Canarias, 38200 La Laguna, Tenerife, Spain}
\altaffiltext{4}{Centre for Astrophysics Research, STRI, University of Hertfordshire, College Lane Campus, Hatfield AL10 9AB, UK}
\altaffiltext{5}{Department of Astrophysics/IMAPP, Radboud University Nijmegen, PO Box 9010, 6500 GL, Nijmegen, The Netherlands}
\altaffiltext{6}{School of Physics \& Astronomy, University of Southampton, Southampton SO17 1BJ, UK}
\altaffiltext{7}{Department of Physics, University of Warwick, Coventry CV4 7AL}
\altaffiltext{8}{Jodrell Bank Center for Astrophysics, School of Physics and Astronomy, University of Manchester, Oxford Street, Manchester, M13 9PL, UK}
\altaffiltext{9}{Astrophysics Research Institute, Liverpool John Moores University, Twelve Quays House, Egerton Wharf, Birkenhead CH41 1LD, UK}
\altaffiltext{10}{Harvard-Smithsonian Center for Astrophysics, 60 Garden Street, Cambridge, MA 02138, USA}
\altaffiltext{11}{Department of Physics, Macquarie University, Sydney 2109, Australia}
\altaffiltext{12}{CASU, Institute of Astronomy, Madingley Road, Cambridge CB3 0HA, UK}
\altaffiltext{13}{Institut f\"ur Physik, Karl-Franzen Universit\"at Graz, Universit\"atsplatz 5, 8010 Graz, Austria}
\altaffiltext{14}{Imperial College London, Blackett Laboratory, Exhibition Road, London SW7 2AZ, UK}
\altaffiltext{15}{Columbia University, Department of Physics, NY 10027, USA}
\altaffiltext{16}{Nordic Optical Telescope, Apartado de Correos 474, 38700 Sta. Cruz de La Palma, Spain}
\altaffiltext{17}{University of Helsinki Observatory, PO Box 14, 00014 University of Helsinki, Finland}
\altaffiltext{18}{Department of Astronomy, University of Cape Town, Rondebosch 7700, Cape Town, South Africa}
\altaffiltext{19}{Department of Physics and Astronomy, University of Sheffield, Sheffield, S3 7RH, UK}

\begin{abstract}

Nova V458 Vul erupted on 2007 August 8th and reached a visual magnitude of 8.1 a few days later. H$\alpha$ images obtained six weeks before the outburst as part of the {\sc IPHAS} galactic plane survey reveal an 18th magnitude progenitor surrounded by an extended nebula. Subsequent images and spectroscopy of the nebula reveal an inner nebular knot increasing rapidly in brightness due to flash ionization by the nova event.  We derive a distance of 13\,kpc based on light travel time considerations, which is supported by two other distance estimation methods.  The nebula has an ionized mass of 0.2\,M$_{\odot}$ and a low expansion velocity: this rules it out as ejecta from a previous nova eruption, and is consistent with it being a $\sim$14,000 year old planetary nebula, probably the product of a prior common envelope (CE) phase of evolution of the binary system.  The large derived distance means that the mass of the erupting WD component of the binary is high.  We identify two possible evolutionary scenarios, in at least one of which the system is massive enough to produce a Type Ia supernova on merging.
\end{abstract}


\section{Introduction}

Classical novae occur when the mass transferred from a Roche-lobe filling
companion onto a white dwarf in a close binary system triggers runaway
thermonuclear burning of hydrogen on the surface of the white dwarf. The
energy released causes the system to brighten by $\sim$10 magnitudes and
ejects $\sim$10$^{-4}$ solar masses of material at speeds of a few hundred
to a few thousand km\,s$^{-1}$ (Prialnik \& Kovetz 1995). It is thought that massive WDs in some
nova systems may eventually produce Type~Ia supernovae (Hillebrandt \& Niemeyer 2000).

Planetary nebulae (PNe) are formed 
when low to intermediate mass stars reach the end of their lives, and 
thermal pulsations driven by extremely temperature-dependent helium 
burning cause the ejection of the stellar envelope during the 
asymptotic giant branch (AGB) phase (Iben \& Renzini 1983). The exposed core, which will become a
white dwarf, then ionizes the ejecta, which moves away from the
star at typical velocities of $\sim$20\,km\,s$^{-1}$. The nebula
eventually fades and merges with the interstellar medium after a few tens
of thousands of years.  It has been proposed that PNe can only be produced by
close binary systems, during their common envelope stage of evolution (Moe \& De Marco 2006).
In this picture, some novae should occur inside PNe but until now the only
known example was GK~Per in 1901 (Bode et al., 1987).

Nova Vul 2007 (V458 Vul) was discovered on 8 August 2007 at magnitude 9.5 
(Nakano et al. 2007). It reached a maximum brightness of V=8.1$\pm$0.1, fading by 
three magnitudes within 21 days, thus being classified as a fast nova. 
Its decline was interrupted by two rebrightenings during the first 20 
days after maximum. In the spectral classification scheme of Williams et al. (1994), V458 Vul has been
classified as a hybrid nova (Poggiani 2008): early post-outburst spectra 
indicated a Fe~{\sc ii} nova while later spectra showed nitrogen lines 
dominating over iron lines, indicating a He/N type.  X-ray emission from the nova has been observed since October 2007, and the nova is currently a super-soft source (Drake et al. 2008), indicating that the nova ejecta have become optically thin, revealing very hot material beneath.

\section{Observations}

The nova lay in a part of the sky which had been imaged six weeks before 
the outburst, on June 27th 2007, in H$\alpha$, $r'$ and $i'$ filters by the 
IPHAS survey (Drew et al. 2005). The IPHAS H$\alpha$ images reveal a 
wasp-waisted nebula surrounding a central star with magnitudes in the Vega 
system of $r'=18.34\pm0.02$, $i'=18.10\pm0.03$ and $H\alpha=18.04\pm0.02$, 
having J2000 coordinates of 19h54m24s.62 and +20$^{\rm o}$52$'$52$''$.2 that 
are coincident within the measurement uncertainties with the position reported
for the nova (Henden \& Munari 2007). 
Figure 1 shows the IPHAS $r'$ and H$\alpha$ images, as
well as a color composite image made from the three IPHAS bands to bring
out the nebular structure. The main nebular ring has a semi-major
axis of $\sim$13.5$''$, corresponding to a physical length of 0.065$D$
pc, where $D$ is the distance to the nebula in kpc. A bright nebular knot
4.5$''$ to the southeast of the central star is prominent in the IPHAS
H$\alpha$ image and absent in the $r'$ and $i'$ images.

\begin{figure}
\begin{center}
\includegraphics[width=0.47\textwidth]{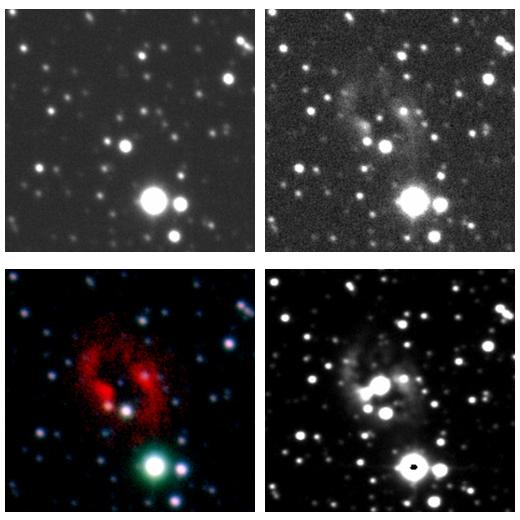}
\end{center}
\caption[]{Isaac Newton Telescope WFC images of the field of V458~Vul.
Top-left: June 27th 2007 IPHAS $r'$ image. Top-right: June 27th 2007
IPHAS H$\alpha$ image, which reveals the nebula. Lower-left: color composite image made
from the IPHAS $r'$, $i'$ and H$\alpha$ images. Lower-right:
May 20th 2008 INT WFC H$\alpha$ image. Each image is 70$''$ on a side.
North is up and East is to the left. The faint central star visible in the
June 2007 images
was in its post-nova decline phase by the time of the May 2008 image,
while the inner edge of the knot to the SE of the central star has brightened.
}
\end{figure}

In September 2007 we obtained intermediate resolution long slit
optical spectra of the nebula from the 2.5m Nordic Optical Telescope (NOT), 
and high resolution optical echelle spectra from the 6.5m Magellan-Clay 
telescope in Chile.  A number of narrow nebular lines were detected in 
our spectra, and their intensities are listed in Table~1, along with line intensities predicted by a photoionization model (discussed below).

H$\alpha$ images were obtained during May 2008 with the NOT and the INT, and 
showed the SE knot to be increasing rapidly in brightness; the May
20th 2008 INT image is shown in Figure~1, below the June 27th 2007
pre-outburst INT image. Figure~2 shows our May 12th 2008 NOT H$\alpha$
image, obtained in excellent seeing conditions.

\begin{table}
\centering
\caption{Observed and modelled relative intensities of emission lines from the south-east knot of the nebula
surrounding V458 Vul, normalised so that I(H$\beta$)=100.  Observed intensities have been dereddened by E(B-V)=0.63 using a Galactic reddening law (Howarth 1983)}.
\label{neblines}
\begin{tabular}{l|l|r|r|r}
\hline
 & &  \multicolumn{1}{c}{{\small NOT}} &
\multicolumn{1}{c}{{\small Magellan}} & \\
 & & \multicolumn{1}{c}{{\small Sep 3/4 2007}} & \multicolumn{1}{c}{{\small Sep 18 2007}} & {\small {\it Model}}$^4$ \\
\hline
$\lambda$(\AA ) & Species & I$_{\lambda}$ & I$_{\lambda}$ & I$_{\lambda}$ \\
\hline
4959           & [O~{\sc iii}]& 245 &      &  {\it 306} \\
5007           & [O~{\sc iii}]& 857 &      &  {\it 913} \\
6548           & [N~{\sc ii}] & 171 &  141 &  {\it 137} \\
6563           & H$\alpha$    & 285 &  285 &  {\it 286} \\
6584           & [N~{\sc ii}] & 504 &  428 &  {\it 419} \\
6716           & [S~{\sc ii}] & 90  &   79 &  {\it 58}  \\
6731           & [S~{\sc ii}] & 90  &   62 &  {\it 61}  \\
\hline
\end{tabular}
\end{table}

\section{Analysis}

The F(H$\alpha$)/F(H$\beta$) ratio of 5.34 measured from a May 14th 2008
spectrum of the SE knot corresponds to a reddening by interstellar dust of
E(B-V)=0.63 magnitudes, assuming an intrinsic H$\alpha$/H$\beta$ intensity 
ratio of 2.72 for the derived nebular physical conditions.
For comparison, the maximum value of E(B-V) predicted for the Galactic
sightline towards V458~Vul ($\ell = 58.63^\circ, b = -3.62^\circ$) is 0.61
(Schlegel et al. 1998). Lynch et al. (2007) derived E(B-V) = 0.6 from
infrared observations of nova O~{\sc i} lines in 2007 October. 
For our subsequent analysis we adopt E(B-V)=0.63, equivalent to 
a visual extinction A$_{\rm V}$ of 1.95 magnitudes.

A crucial parameter in understanding V458 Vul is its distance.  
The distance to novae is often estimated using maximum magnitude versus
rate of decline (MMRD) relations (Warner 1995). 
Taking V=8.1 at maximum light from the AAVSO light curve (see
www.aavso.org) and a time to decline 3 magnitudes from peak, $t_3$, of 21
days, then for A$_{\rm V} = 1.95$, a `super-Eddington' MMRD (Downes \& D\"urbeck 2000) 
gives D = 11.6 kpc.  Using a $t_2$ relationship from Downes \& D\"urbeck (2000) with
$t_2 = 8$ days gives D = 13.5 kpc. The unusual form of the light curve
makes the definition of $t_2$ and $t_3$ difficult in this case.
Alternatively, one can use the $t_{15}$ relationship (Downes \& D\"urbeck 2000), where 
the absolute magnitudes of classical novae are assumed to be
the same at 15 days post-maximum. This gives D = 10.0 kpc.

Another estimate for the distance to nova V458 Vul can be obtained by
assuming that our measured mean radial velocity of $v_{\rm LSR} =
-60.6\pm4.3$~km~s$^{-1}$ for its nebula is due to Galactic
rotation alone. If we adopt a distance to the Galactic Centre of
7.6$\pm$0.3~kpc (Eisenhauer et al. 2005) and 220~km~s$^{-1}$ for the
Galactic rotation speed, independent of Galactic radius (Brand \& Blitz 1993), 
a distance of $13.1\pm0.5$~kpc from the Sun is obtained.
Allowing that up to 30~km~s$^{-1}$ of $v_{\rm LSR}$ may be due to random
motion (Veltz et al. 2008), the corresponding distance uncertainty would
be $\pm3$~kpc.

Finally, we estimate the distance to the nova from the time taken 
for the nova flash to reach the SE nebular knot - the first time 
it has been possible to determine the distance to a nova in this way. The
pre-outburst June 27th 2007 IPHAS WFC H$\alpha$ image (obtained in
1.4$''$ seeing) showed the peak brightness of the SE knot to lie
4.5$\pm$0.3$''$ from the central star (CS). Our May 20th 2008 WFC
H$\alpha$ image, obtained in 1.6$''$ seeing 286 days after the nova
outburst (Figure~1), shows the region of peak emission from the SE knot to
have significantly increased in brightness and to have moved closer to the
CS, peaking 3.5$\pm$0.3$''$ from the CS. A NOT H$\alpha$ image obtained
in 0.57$''$ seeing on May 12th 2008 (Figure~2), shows the peak emission
from the knot to lie $3.6\pm0.1$$''$ from the CS. For the latter
separation, the light travel time of 278 days from the nova to the knot
implies a distance of 13.4~kpc to V458~Vul, if the line joining the knot
and nova lies in the plane of the sky. The radial velocities of the SE
knot and its fainter counterpart on the opposite side of the nova differ
by only $3.0\pm2.5$~km~s$^{-1}$, indicating that their motions do indeed lie close
to the plane of the sky. Allowance for the line joining nova and knot
being up to $\pm30^{\rm o}$ out of the plane of the sky would give a
corresponding distance uncertainty of $\pm2$~kpc.

\begin{figure}
\begin{center}
\includegraphics[width=0.42\textwidth]{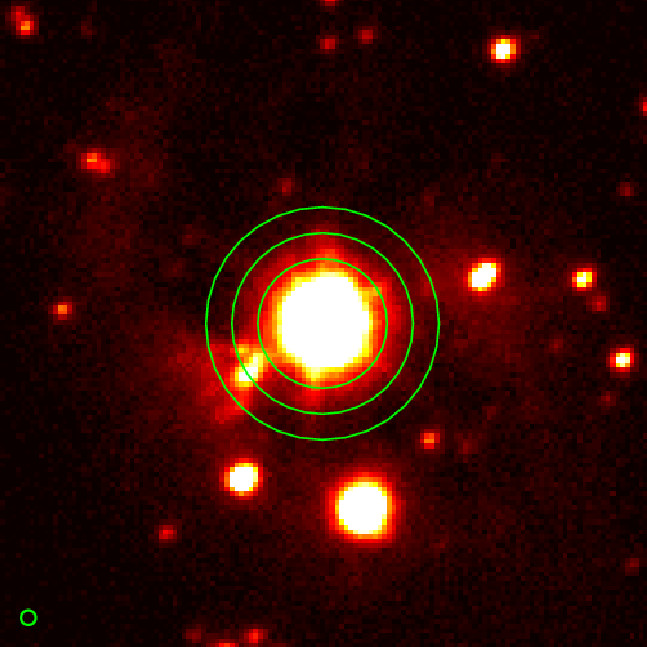}
\end{center}
\caption[]{H$\alpha$ image of the V458~Vul nebula obtained
on May 12th 2008 in 0.57~arsec seeing, 278 days after the nova outburst,
using ALFOSC on the 2.5m Nordic Optical Telescope. The
superposed circles centred on the nova system have angular
radii of 2.5, 3.5 and 4.5". The circle at the lower left
corresponds to the seeing FWHM of 0.57".
}
\end{figure}

We adopt a distance of 13~kpc to nova V458 Vul, corresponding to a
Galactocentric distance of 11~kpc and to a depth of 0.8~kpc below the
mid-plane of the Galactic disk. Fast novae are concentrated toward the Galactic plane ($z <
100$ pc) (Della Valle et al. 1992), so V458~Vul appears to be unusually situated in the Galaxy given
its class.  The absolute B magnitude of V458~Vul at maximum ($-9.3$) corresponds to a thermonuclear
runaway on a white dwarf of mass 1.3~M$_{\odot}$ (Livio 1992).

The emission lines from the spatially resolved nebula surrounding V458~Vul
are much narrower than those typically found in old nova shells. From our
September 2007 Magellan echelle spectra (9.2~km~s$^{-1}$ instrumental
resolution), we measure a full-width at half-maximum of
13.7$\pm$1.0~km~s$^{-1}$ for the [N~{\sc ii}] lines at several nebular positions, fully two orders of magnitude smaller than the line widths normally associated with nova ejecta.

We measured the total H$\alpha$ flux emitted by the nebula prior to
the nova event from the IPHAS H$\alpha$ image, correcting for the
contribution of foreground stars in the field and for [N~{\sc ii}]
emission caught by the $\sim$95{\AA}-wide filter. We obtain a flux of
(1.0$\pm$0.25)$\times$10$^{-13}$~erg\,cm$^{-2}$\,s$^{-1}$. Correcting for
our measured interstellar extinction gives a dereddened H$\alpha$ flux of
(4.3$\pm$1.1)$\times$10$^{-13}$ erg\,cm$^{-2}$\,s$^{-1}$.  From the 
[S~{\sc ii}] line ratio in the Magellan spectra, we infer a mean 
electron density of 155\,cm$^{-3}$. Adopting N(He)/N(H)=0.11, the 
total nebular ionized mass implied by this H$\alpha$ flux is
0.2\,M$_{\odot}$. Novae
typically eject $\sim$10$^{-4}$~M$_{\odot}$ of material (Yaron et al. 2005). 
The symmetry of the nebula surrounding Nova Vul and its wasp-waisted shape 
strongly suggest that it is not simply ionised interstellar medium.  
Instead, it is likely to be an old planetary nebula.  Its 
low expansion velocity and large ionised mass both indicate a slow ejection of 
material by a red giant, and are in the range observed for planetary 
nebulae (Gussie \& Taylor 1994, Barlow 1987, Boffi \& Stanghellini 1994), 
while both its bipolar morphology and spectral characteristics of strong 
[N~{\sc ii}] and [S~{\sc ii}] emission relative to H$\alpha$ indicate 
that it is a Type~I planetary nebula, originating from a relatively 
high-mass progenitor (Torres-Peimbert \& Peimbert 1997).

The centroid of the SE bright knot, 4.5$''$ from the central star 
in June 2007, corresponds to a projected distance from the central star 
of 0.28 pc. The adoption of a median planetary nebula expansion velocity of 
20 km~s$^{-1}$ (Gussie \& Taylor 1994) would imply that the SE knot was ejected 
about 14,000 years ago. Parts of the main nebular ring have projected 
separations from the central star that are up to three times larger, 
implying significantly larger expansion ages unless their expansion 
velocities are also proportionately larger. Such self-similar expansion 
is generally observed in planetary nebulae (Corradi 2004).

\section{Discussion}

The first object to be described as a planetary nebula around a nova was
GK~Per, which erupted in 1901 and was later found to be surrounded by a
very large ($>40$~arcmin) bipolar-shaped region of dust and ionized gas,
interpreted as a 10$^5$ year old massive planetary nebula (Bode et al. 1987, 2004). 
V605~Aql was not of this type, even though its 1919 eruption
occurred inside the 70$''$ diameter planetary nebula A~58:
instead, its slow outburst and relatively low velocity ejecta fit better 
to a very late thermal pulse rather than to a classical nova event (Pollacco et al. 1992). 
The high ejection velocities (1500-2000~km~s$^{-1}$), fast speed class and
outburst amplitude ($\sim$10 magnitudes) of V458 Vul's 2007 eruption are all
consistent with its classification as a classical nova.
 
The observed IPHAS $r'-i'$ and $r'-H\alpha$ colors of V458~Vul before the
nova event ($0.245\pm0.023$ and $0.305\pm0.018$, respectively) 
deredden to match those of a hot O-type star with an H$\alpha$ emission 
line equivalent width of 7-10~\AA\ (for $E(B-V)=0.63$). Its
dereddened $r'$ magnitude of 16.69 corresponds to an $r'$ absolute
magnitude of +1.1, approximately 2.5 magnitudes brighter than typical
absolute magnitudes for classical nova systems in quiescence (Warner 1995). 
However, 5 out of 11 novae with good pre-eruption coverage showed slow 
rises of 0.3-1.5 mag in the 1-15 yrs before eruption (Warner 1995).
The USNO A2.0 and B1.0 catalogs give R magnitudes for V458 Vul of 18.1 in
1950.542 and 17.87 in 1975.1, while the STScI GSC2.2 catalog gives an R
magnitude of 18.31 in 1991.688, offering no clear evidence of brightening
during the 60 years leading up to its August 2007 eruption.

Using the 3D photoionization code {\sc mocassin} (Ercolano et al 2003), we constructed models of 
the SE knot.  We find that the pre-flash line intensity ratios can be well matched by a model in which the knot is inhomogeneous, with an electron density in the hemisphere nearest the star of 800\,cm$^{-3}$ and in the far hemisphere of 50\,cm$^{-3}$.  The modelled knot abundances were solar for oxygen, 3$\times$ solar for nitrogen and 1.5$\times$ solar for sulphur, and the central source yielding the best match to the observed intensities had a temperature T$_{\rm
eff}=90,000$~K, a radius $R_*$ of 0.23~$R_{\odot}$ and a luminosity $L_* =
3000~L_{\odot}$.  For H-burning central star evolutionary tracks
(Vassiliadis \& Wood 1994), this corresponds to a core mass of
0.58~M$_{\odot}$ and an age since leaving the AGB consistent with
a nebular expansion age of $\sim14,000$~yrs.

Standard models for classical novae envisage the transfer of material from
a main sequence or slightly evolved red dwarf star onto the surface of a degenerate white
dwarf (WD), leading to a thermonuclear runaway at infrequent intervals.
The GK~Per system has a K1~IV secondary (Morales-Rueda et al. 2002) and it has been proposed 
that its massive extended nebula originated from this evolved secondary star 
during a common envelope phase.  Roche Lobe overflow onto
the white dwarf with sufficiently high mass transfer rates then converted 
the WD into a `born-again' AGB star which lost material at low velocities, 
producing over 10$^5$~yrs the 40$'$ nebula seen around GK~Per today (Dougherty et al. 1996). 
When the transfer rate dropped below
$10^{-7}$~M$_{\odot}$~yr$^{-1}$, the matter accreting onto the WD
eventually produced a nova outburst, the first being in 1901. A similar
scenario could be invoked for V458~Vul, in which case the high luminosity
hot central source needed to maintain the ionization of its nebula (which
would otherwise recombine in less than a thousand years) must come from
the accretion disk around the WD.
A 1.3\,M$_{\odot}$ WD of radius 2.9$\times$10$^8$ cm accreting at 
$\sim10^{-7}$~M$_{\odot}$~yr$^{-1}$ should produce an accretion luminosity
of a few thousand $L_{\odot}$, similar to that required. The maximum disk 
temperature would then be $\sim$270,000\,K, and the temperature corresponding 
to the luminosity-weighted average disk radius would be comparable to the 
$\sim$90,000\,K required to maintain the observed degree of nebular ionization.

An alternative scenario for V458~Vul is one whereby the star that
dominated the pre-outburst light was a 90,000~K 0.6~M$_{\odot}$ PN central
star, as deduced from our photoionisation model, with the nova outburst
taking place on a 1.3~M$_{\odot}$ WD companion. The
nebula would also be the product of a common envelope phase. 
In this scenario it is unclear if the nova mass can be 
accreted after the common envelope phase, as the rapidly shrinking radius of 
the PN central star during its post-AGB evolution would make it difficult to 
maintain contact with its Roche lobe and continue mass transfer. More likely, 
and more interesting, is the possibility that the material was accreted before, 
or even during the common envelope phase.  In this scenario, the 
combined mass of the two (proto) white dwarfs is above the 
Chandrasekhar limit and if the current orbital period is below 0.5 days, 
the system will merge within a Hubble time and is a potential Type~Ia 
supernova progenitor.

Until V458 Vul has been studied much more thoroughly, these ideas about the prior evolution of the binary are necessarily schematic.  Modelling of the common 
envelope phase proposed under each of the above scenarios could help 
elucidate whether either model can account for all of  V458~Vul's unusual 
characteristics, while a determination of an accurate orbital period for 
the post-nova system would provide crucial additional information.  But it is
already clear that V458 Vul and its nebula have the potential to become a
very important challenge to our understanding of the common envelope phase in close binary evolution.

\acknowledgements
This paper makes use of data obtained as part of the INT Photometric
H$\alpha$ Survey of the Northern Galactic Plane (IPHAS), carried out by
the Isaac Newton Telescope at La Palma Observatory. All IPHAS data are
processed by the Cambridge Astronomical Survey Unit, at the Institute of
Astronomy in Cambridge.  DS acknowledges an STFC Advanced Fellowship 

{\it Facilities:} \facility{ING:Newton (WFC)}, \facility{NOT (ALFOSC)}, \facility{Magellan:Clay (MIKE)}, \facility{ING:Herschel (ISIS)}.  We gratefully acknowledge use of these facilities.

\end{document}